\PassOptionsToPackage{dvipsnames,table}{xcolor}
\documentclass[screen,sigconf,nonacm]{acmart}
\setcopyright{none}

\AtBeginDocument{%
  \providecommand\BibTeX{{%
    \normalfont B\kern-0.5em{\scshape i\kern-0.25em b}\kern-0.8em\TeX}}}

\usepackage[dvipsnames]{xcolor}
\usepackage{comment}
\usepackage{multirow}
\usepackage{subcaption}
\usepackage{enumitem}

\usepackage{fbox}
\usepackage{tikz}
\newcommand\highlight[2][]{\tikz[overlay]\node[inner sep=2pt, anchor=text, rectangle, rounded corners=1mm,#1] {#2};\phantom{#2}}

\newcommand{\medicine}{\,\highlight[fill=CornflowerBlue!40]{\texttt{MED}}\,}
\newcommand{\journalism}{\,\highlight[fill=WildStrawberry!40]{\texttt{JRN}}\,}
\newcommand{\welfare}{\,\highlight[fill=BurntOrange!40]{\texttt{PHW}}\,}
\newcommand{\legal}{\,\highlight[fill=LimeGreen!40]{\texttt{LEG}}\,}

\newcommand{\med}[1]{\,\highlight[fill=CornflowerBlue!20]{{#1}}\,}
\newcommand{\jrn}[1]{\,\highlight[fill=WildStrawberry!15]{{#1}}\,}
\newcommand{\phw}[1]{\,\highlight[fill=Apricot!25]{{#1}}\,}
\newcommand{\leg}[1]{\,\highlight[fill=LimeGreen!20]{{#1}}\,}

\newcommand{\qq}[1]{\emph{``{#1}''}}
\newcommand{\qqq}[2]{\begin{quotation}\emph{``{#1}''} {(#2)}\end{quotation}}

\begin{document}

\title{Why Do Decision Makers (Not) Use AI? A Cross-Domain Analysis of Factors Impacting AI Adoption}

\author{Rebecca Yu}
\affiliation{%
    \institution{Carnegie Mellon University}
    \city{Pittsburgh, PA}
    \country{United States}}
    \email{rebyu@cs.cmu.edu}

\author{Valerie Chen}
\affiliation{%
    \institution{Carnegie Mellon University}
    \city{Pittsburgh, PA}
    \country{United States}}
    \email{valeriechen@cmu.edu}

\author{Ameet Talwalkar}
\affiliation{%
    \institution{Carnegie Mellon University}
    \city{Pittsburgh, PA}
    \country{United States}}
    \email{talwalkar@cmu.edu}

\author{Hoda Heidari}
\affiliation{%
    \institution{Carnegie Mellon University}
    \city{Pittsburgh, PA}
    \country{United States}}
    \email{hheidari@cmu.edu}

\renewcommand{\shortauthors}{Yu, et al.}
\renewcommand{\shorttitle}{Why Do Decision Makers (Not) Use AI?}

\begin{abstract}
Growing excitement around deploying AI across various domains calls for a careful assessment of how human decision-makers interact with AI-powered systems.
In particular, it is essential to understand when decision-makers voluntarily choose to consult AI tools, which we term \emph{decision-maker adoption}. 
We interviewed experts across four domains---medicine, law, journalism, and the public sector---to explore current AI use cases and perceptions of adoption. 
From these interviews, we identify key factors that shape decision-maker adoption of AI tools: the decision-maker's background, perceptions of the AI, consequences for the decision-maker, and perceived implications for other stakeholders. 
We translate these factors into an AI adoption sheet to analyze how decision-makers approach adoption choices through comparative, cross-domain case studies, highlighting how our factors help explain inter-domain differences in adoption.
Our findings offer practical guidance for supporting the responsible and context-aware deployment of AI by better accounting for the decision-maker's perspective.
\end{abstract}

\maketitle

\section{Introduction}

Increased interest in deploying practical AI tools for decision-makers has spurred growing research into how human decision-makers interact with these tools, particularly regarding decision-maker reliance on AI~\citep{lee2020human, bansal2021does, chen2023understanding}---that is, the rate at which a human decision-maker replaces their initial judgments with AI recommendations.
Such studies typically involve controlled evaluations of AI prototypes in lab settings~\citep{cai2019hello, silva2024leveraging} or pilot deployments of AI tools in real-world environments~\citep{jo2023understanding, kawakami2022improving}, where decision-makers are presented with an AI tool and researchers measure the extent to which they defer to AI guidance.
However, AI reliance presupposes decision-maker adoption; in practice, decision-makers may opt not to use these tools at all (\emph{e.g.}, a doctor may not use a model without peer endorsement~\citep{henry2022human}).

\begin{figure}[t]
    \centering
    \fcolorbox{black}{white}{
    \begin{minipage}[t]{0.95\columnwidth}
        \centering \textbf{AI Adoption Sheet}
        \begin{itemize}[left=5pt]
            \item \textbf{Decision-maker background.} 
            \begin{itemize}[left=3pt]
                \item How does a decision-maker's professional experience inform their adoption behaviors?
                \item Does the decision-maker have any personal biases that may impact their adoption choices?
            \end{itemize}
            \item \textbf{Perceptions of the AI model.} 
            \begin{itemize}[left=3pt]
                \item What flaws does the decision-maker perceive the model they are deciding to consult to have? 
                \item Do they have any doubts about model trustworthiness or capabilities?
                \item What do they believe the capabilities of the model and the benefits of AI adoption are?
            \end{itemize}
            \item \textbf{Consequences for the decision-maker.} 
            \begin{itemize}[left=3pt]
                \item What liabilities may the decision-maker face (\emph{e.g.,} legal, professional, political)? 
                \item Are there uncertainties regarding consequences?
                \item How does the decision-maker's workload change?
            \end{itemize}
            \item \textbf{Perceived implications for other stakeholders.} 
            \begin{itemize}[left=3pt]
                \item How do decision-makers believe organizations are impacted by AI adoption?
                \item How do decision-makers believe decision-subjects are impacted by AI adoption?
            \end{itemize}
        \end{itemize}
    \end{minipage}
    }    
    \caption{
    We introduce an AI adoption sheet to highlight factors influencing decision-maker adoption.
    }
    \label{fig:checklist}
\end{figure}

\begin{figure*}[t]
      \centering
    \includegraphics[width=\linewidth]{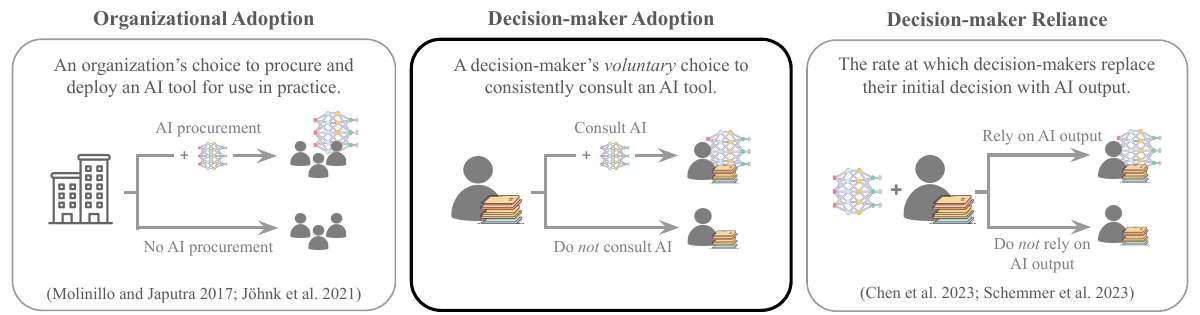}
    \caption{Key definitions to distinguish between organizational adoption, decision-maker adoption, and decision-maker reliance. Adoption refers to an organization or decision-maker's \emph{choice} to procure or consult an AI tool. In contrast, decision-maker reliance is the rate at which decision-makers replace their initial decision with AI output, which presupposes adoption.}
  \label{fig:definitions}
\end{figure*}
As such, we emphasize the importance of studying \emph{decision-maker adoption} of AI---the voluntary and consistent use of an AI tool within a decision-maker's workflow.
Prior work on AI adoption has primarily focused on the perspective of \emph{organizational adoption choices}~\citep{sun2019mapping, chatterjee2021understanding, wu2024characterizing, neumann2024exploring}, namely, why an organization may choose to adopt an AI product or service.
For example, such work may involve surveys to investigate why a hospital might choose to procure and deploy an AI tool~\citep{poon2025adoption}, potentially irrespective of whether it is used by clinicians down the line.
We build on this line of research, but shift the focus to the perspective of \emph{decision-makers} within organizations that have already made AI available to their members.
While a few prior studies have taken a more decision-maker-centric approach in specific domains---primarily medicine~\citep{hameed2023breaking, fang2023ai, bacsar2024factors}, our goal is to identify cross-cutting factors that affect such decision-makers' adoption of AI tools.

In this work, we conduct a cross-domain analysis to identify inter-domain trends of decision-maker AI adoption choices.
We select four representative domains (journalism~\citep{verma2024impact}, law~\citep{contini2024unboxing}, medicine~\citep{park2019identifying}, and the public sector~\citep{amiri2022chatbot}) to cover a variety of predictive and generative AI applications across different sectors of the economy, with varying stakes, and include decision-makers from diverse backgrounds. In total, we conducted 16 semi-structured interviews with decision-makers on their familiarity with AI tools and their considerations for adoption.

Through our interviews, we identify four common key factors that impact decision-maker adoption across domains: (1) the decision-maker's background, (2) their perceptions of the AI model, (3) consequences for the decision-maker, and (4) the perceived implication of AI adoption for other stakeholders. 
Through the lens of these four factors, we examine why decision-maker adoption choices for similar AI use cases vary across domains. In particular, we provide two case studies of AI use cases surfaced by participants.
First, we discuss why AI is often adopted as an \emph{e-discovery} tool to help gather, organize, and analyze relevant digital data in journalism, law, and the public sector, but not necessarily in medical contexts. 
Then, we explore why decision-makers in law choose not to use AI for \emph{tailoring communications}, while decision-makers in other domains do.
These case studies highlight inter-domain differences in adoption behavior and demonstrate how our proposed four-factor framework can explain when decision-makers will (not) adopt AI.

We envision our framework to serve as a guide to facilitate the proactive assessment of AI proposals for uptake in practice. 
To provide a structured guide to help researchers, model developers, and model designers think more critically about the key adoption considerations of decision-makers, we turn our factors into an \emph{AI adoption sheet} to help facilitate the integration of decision-maker perspectives into the model development process (Figure~\ref{fig:checklist}).
This sheet can serve as a starting point for organizations and policymakers to develop a comprehensive understanding of how decision-makers' AI adoption choices may manifest in practice. We leave the empirical assessment of the efficacy and usability of the proposed sheet as a critical direction for future work.
\section{Related Work}

As AI-powered systems are increasingly considered for real-world workflows, research on how decision-makers interact with AI has grown rapidly. 
We begin by reviewing prior work on \emph{decision-maker reliance}, as it is the most frequently studied aspect of these interactions. However, for decision-makers to rely on AI, they must first choose to consult an AI tool regularly---which we refer to as their \textit{adoption choice}. As such, we then discuss work on AI adoption, highlighting the difference between organizational and decision-maker adoption choices (Figure~\ref{fig:definitions}).

\begin{table*}
    \fontsize{10}{10}
    \centering
    \begin{tabular}{*6c}
         \toprule
         \multirow{2}{*}{\textbf{Domain}}   &   \multirow{2}{*}{\textbf{Alias}}   &   \textbf{Domain Experience}  & \textbf{AI} &   \textbf{AI Interaction} & \textbf{Primary Area}\\
         &  &  \textbf{(Years)} &  \textbf{Familiarity} &  \textbf{Frequency}   &   \textbf{of Expertise}   \\
         \midrule
         \multirow{4}{*}{\journalism}
         & \jrn{P01} & 4      & Expert        & Daily   &   Research\\
         & \jrn{P02} & 20+    & Beginner      & Occasionally   &   Domain\\[0.3mm]
         & \jrn{P03} & 20+    & Intermediate  & Frequently   &   Domain\\[0.3mm]
         & \jrn{P04} & 10-19  & Advanced      & Occasionally   &   Domain\\[0.3mm]
         \midrule
         \multirow{2}{*}{\legal}
         & \leg{P05} & 20+    & Intermediate  & Frequently   &   Domain\\[0.3mm]
         & \leg{P06} & 10-19  & Advanced      & Daily   &   Research\\[0.3mm]
         \midrule
         \multirow{5}{*}{\medicine}
         & \med{P07} & 20+    & Intermediate  & Daily   &   Domain\\[0.3mm]
         & \med{P08} & 10-19  & Advanced      & Frequently   &   Domain\\[0.3mm]
         & \med{P09} & 5-9    & Advanced      & Daily   &   Domain\\[0.3mm]
         & \med{P10} & 10-19  & Expert        & Daily   &   Research\\[0.3mm]
         & \med{P11} & 10-19  & Intermediate  & Occasionally   &   Domain\\[0.3mm]
         & \med{P12} & 20+    & Intermediate  & Occasionally   &   Domain\\[0.3mm]
         \midrule
         \multirow{4}{*}{\welfare}
         & \phw{P13} & 10-19  & Expert        & Daily   &   Research\\[0.3mm]
         & \phw{P14} & 10-19  & Expert        & Daily   &   Research\\[0.3mm]
         & \phw{P15} & 10-19  & Beginner      & Daily   &   Domain\\[0.3mm]
         & \phw{P16} & 20+    & Intermediate  & Occasionally   &   Domain\\[0.3mm]
         \bottomrule
    \end{tabular}
    
    \caption{Summary of interview participants along with their relevant domain experience and comfort level with AI (Beginner = Some basic understanding, but limited knowledge, Intermediate = Solid foundational knowledge, Advanced = Deep understanding of core concepts, Expert= Extensive knowledge and experience).}
    \label{table:interviewees}
\end{table*}

\subsection{Decision-maker Reliance on AI}\label{sec:rw:reliance}

A key goal of AI-assisted decision-making is to foster \emph{appropriate reliance}: ensuring that decision-makers can discern when to rely on AI recommendations and when to rely on their own expertise~\citep{schemmer2023appropriate}. In practice, simply optimizing for algorithmic accuracy does not guarantee appropriate reliance choices by decision-makers~\citep{cai2019hello}. Empirical studies have documented patterns of both underreliance (also termed algorithmic aversion)~\citep{dietvorst2015algorithm, logg2019algorithm, cai2019hello} and overreliance, where users accept flawed AI outputs~\citep{bansal2019beyond, passi2022overreliance}. These patterns have motivated work exploring how and when AI outputs influence human decisions~\citep{reverberi2022experimental, schelble2024towards, schoeffer2024explanations}.

To mitigate misaligned reliance choices, researchers have proposed interventions such as providing more comprehensive information on model capabilities and limitations~\citep{cai2019hello, yang2019unremarkable, thieme2023designing}. More recently, studies have moved beyond controlled settings to examine decision-maker reliance choices in real-world deployments of AI tools~\citep{kawakami2022improving, jo2023understanding}. Further, as researchers study decision-maker reliance choices in real-world settings, they have highlighted an adoption gap, where the actual adoption and use of models lag behind their development~\citep{dwivedi2021artificial, sethi2020interpretable, uren2023technology, cubric2020drivers}.
Our work aims to uncover the underlying causes of such lags from the perspective of decision-makers.

\subsection{Adoption of AI in Practice}\label{sec:rw:org}

\textbf{Organizational adoption.} 
Prior work on AI adoption has primarily focused on \emph{organizational} adoption, examining how factors such as technology readiness and innovation culture shape adoption choices~\citep{harvey2021regulatory, uren2023technology, campion2022overcoming}. 
Through a combination of surveys, interview studies, and literature reviews on AI deployment, researchers have explored the drivers and challenges of organizational adoption in various domains---including medicine~\citep{davenport2022factors}, business and management~\citep{cubric2020drivers}, construction~\citep{regona2022opportunities}, and public administration~\citep{madan2023ai}). However, findings across these studies can be fragmented, as each study typically focuses on a single distinct domain.
We include a brief survey of these factors in the Appendix. 

Across ten representative papers~\citep{cubric2020drivers, harvey2021regulatory, regona2022opportunities, bedue2022can, campion2022overcoming, madan2023ai, merhi2023evaluation, uren2023technology, neumann2024exploring, polisetty2024determines}, recurring factors include: 
issues with data quality or availability (10/10);
lack of appropriate data governance and regulation (5/10);
and organizational goals to optimize cost or time savings (4/10). 
While these organizational factors shape the deployment of AI tools and understanding such organizational adoption choices is important, the final choice to use AI tools rests with the individual \emph{decision-makers} responsible for integrating these tools into daily workflows. These decision-makers may choose \emph{not} to use AI, even in the face of organizational support or pressure~\citep{kawakami2024studying}. We will revisit these factors in Section~\ref{sec:discussion:comparison}.

\textbf{Decision-maker adoption.} 
Far fewer studies directly address adoption from the perspective of the decision-maker, and those that do often focus on a single domain or use case.
For example, \citet{hameed2023breaking} determined the intentions of healthcare professionals to adopt new technologies through a cross-sectional survey. 
They predefined seven constructs, which they evaluated to either directly or indirectly influence behavioral intentions. These factors are: performance expectancy, effort expectancy, personal innovativeness in information technology, task complexity, technological characteristics, initial trust, and social influence. 
In legal use cases, \citet{fang2023ai} leveraged semi-structured interviews to understand the attitudes of legal professionals towards AI and identify factors that influence how they engage with AI tools. Their findings revealed that legal professionals and semi-professionals often hold contradictory attitudes towards AI tools. 
We build on and extend this prior work by looking at AI adoption across multiple domains from the perspective of decision-makers. 
By identifying cross-domain factors, we aim to draw broader conclusions that can inform future AI deployments in new domains. 
In Section~\ref{sec:discussion:comparison}, we contrast our findings with factors identified in prior work to show why decision-maker perspectives and a cross-domain analysis are essential for understanding real-world AI adoption choices.
\section{Study Design and Methodology}

To understand what factors influence decision-makers' adoption choices, we conducted an interview study with 16 individuals across four professional domains. Below, we describe our participant recruitment process, study procedures, and the analysis approaches used to synthesize interview data. Participation in the study was voluntary, and the protocol was IRB-approved. We provide further details on our protocol in the Appendix.

\subsection{Participants}
\label{sec:methods:domains} 

We focus on four domains: journalism (\journalism), the legal profession (\legal), medicine (\medicine), and the public sector (\welfare)---encompassing both public health and welfare.
Throughout this paper, we color-code these domains to enhance clarity and facilitate cross-domain comparisons.
These domains were selected based on three criteria. 
First, each domain has prior research on AI applications and adoption, enabling participants to speak about concrete, real-world use (or disuse) of AI tools.
Second, the domains together must represent a broad spectrum of AI applications, spanning both traditional predictive models and newer generative systems. 
Third, domains must include participants working in varied professional contexts to capture a breadth of decision-makers.
We group ``public health'' and ``welfare'' into a composite domain (\welfare), as both operate on population-level interventions and share similar methodological approaches.

From each domain, we recruited individuals in the United States who fell into one of two categories: (1) practitioners with professional experience in one of the domains and direct engagement with AI tools, and (2) researcher-insiders who have worked closely and extensively with such practitioners.
Including the second category of participants helped mitigate recruitment challenges by broadening the pool of eligible participants while ensuring that all participants could speak meaningfully to the real-world use (or disuse) of AI by decision-makers. 
 
To recruit participants, we began by compiling a list of potential candidates, identified through collaborators in each domain and by locating individuals who had published articles---whether research or opinion---about their experience with AI in their respective domains. 
We then applied snowball sampling~\citep{goodman1961snowball, parker2019snowball} to expand our reach and identify additional candidates.
We contacted a minimum of 10 individuals within each domain.
Despite our effort to diversify interviews across domains, we found that response rates varied substantially: we received 4 responses out of 13 contacts from \welfare, 6 out of 24 for \medicine, 2 out of 11 for \legal, and 4 out of 10 from \journalism.

\subsection{Semi-structured interviews}
We conducted 16 semi-structured interviews led either by the first and second author (n=7) or one-on-one with the first author (n=9). The interviews were scheduled via email, and participants completed a consent form before the interview.
All interviews were conducted over Zoom and lasted between 45-60 minutes. 
Participants received a \$100 gift card as compensation.

At the start of each interview, we asked participants to describe their current role and domain experience. 
We also explored their level of familiarity with AI and the extent of their interaction with AI tools within their domain.
Participants were then invited to discuss AI use cases they had either encountered or were otherwise familiar with. 
Questions were intentionally left open-ended, encouraging discussion of a range of AI use cases---simple or complex. 
For each use case discussed, we asked follow-up questions to better understand how AI tools were applied in practice, how decision-makers interacted with them, and what considerations influenced decision-makers to choose (not) to use AI tools.
Additional details about participant backgrounds can be found in Table~\ref{table:interviewees}.

\subsection{Analysis approach}
We recorded and transcribed all interviews via Zoom in addition to taking observational notes during each session.
Following each interview, we reviewed and edited audio transcriptions for accuracy. 
We conducted open-coding of the interview transcripts and observational notes, followed by a reflexive thematic analysis~\citep{braun2006using, braun2019reflecting, braun2021can}. Through multiple iterations, we developed and refined the set of factors presented in Section~\ref{sec:cf}. 
In parallel, we created codes to capture AI use cases mentioned by participants, detailed in the Appendix, which informed our discussion of prevalent use cases in Section~\ref{sec:case-study}). 
Quotations in this paper have been lightly edited for clarity and readability.

\section{Factors Impacting AI Adoption}\label{sec:cf}

\begin{figure}[tb]
  \centering
  \includegraphics[width=\columnwidth]{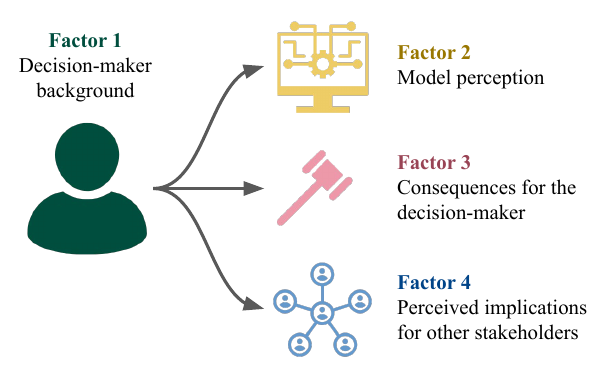}
    \caption{We identified four factors that influence AI adoption choices across domains. All factors are centered around the decision-maker and shape how and why decision-makers choose to adopt AI tools in practice.}
  \label{fig:factors}
\end{figure}

In our cross-domain analysis, we identified four key factors that shape why decision-makers may or may not choose to adopt AI tools (Figure~\ref{fig:factors}).

\subsection{Decision-maker background}\label{sec:cf:dm-background}

The first factor focuses on the decision-makers themselves. 
In particular, a decision-maker's \textbf{professional experience} influences how they engage with AI support tools. 
Across our interviews, we found that decision-makers often prioritize their professional expertise in AI-generated recommendations, especially in contexts requiring nuanced human judgment. 
Several participants echoed \qq{the best clinicians personalize their decisions around the patient} rather than focusing on what may objectively be the best based on various statistics (\emph{e.g.,} AI recommendations) (\med{P08}).
One reason is that years of professional experience can provide additional context not captured by models but critical to decision or task outcomes. 
As \phw{P14} noted in \welfare, recommendations from AI tools are \qq{just another piece of information for decision-makers to consider}, whereas \qq{people are usually more swayed by personal stories, their own interpretations, and their past experiences} (\phw{P14}). 

Beyond professional experience, decision-makers' \textbf{personal biases} may also play a role in adoption choices. 
For example, some decision-makers may prefer tools that support rather than replace them.
\phw{P16} mentioned that while they occasionally use AI tools for brainstorming, \qq{I have a little bit of bias---or perhaps it's how I was schooled---but I do believe in authoring everything I write myself.}

\subsection{Model perception} \label{sec:cf:model-perception}
A decision-maker's perception of AI tools also impacts their adoption choices. 
Participants' concerns clustered around three main areas: concerns about model flaws, perception of model transparency and trustworthiness, and faulty beliefs in model capabilities.

\paragraph{Concerns about model flaws.}
Many participants noted that AI tools often underperform and fall short of expectations. 
For example, in a \welfare\ use case, \phw{P15} described AI-generated social media content as \qq{comparable to an intern level and something to work with and play with}, but not mature enough that decision-makers \qq{would feel comfortable putting out there just yet.}
Concerns about \qq{unexplained phenomena like hallucinations} or opaque black-box mechanisms were widespread amongst participants (\jrn{P02}).
Hallucinations are particularly concerning in public health. \phw{P15} specifically mentioned \qq{We have to be very careful about what we say and how we say it to not feed misinformation, perpetuate it, or misguide people}. 
Some participants described the improper use of AI as \qq{disturbing} (\med{P07}) and alarming to decision-makers (\leg{P05}), noting that decision-makers have become highly cautious to the point that \qq{it is now a little shameful to even just use ChatGPT for legal work} (\leg{P06}). 

However, perceptions can change over time. Participants in all our domains have described increasingly positive perceptions of AI tools with increased exposure to models, especially if there is also organizational adoption and support: \qqq{I think that initially, clinicians were not into AI tools, but I think with the new technologies and interfaces that now exist, I'm starting to see clinicians trying it out and being more interested. [Institution] recently launched a secure form of a language model hospital-wide for clinicians to try, and I know people are trying that out.}{\med{P09}}

\paragraph{Perception of model transparency and trustworthiness.} \label{sec:cf:model-perception:transparency-trustworthiness}
Several participants mentioned that model decisions are significantly more opaque than human decisions, with the lack of explanations making mistakes harder to identify (\jrn{P01}, \leg{P05}, \leg{P06}, \med{P08}, \med{P09}, \phw{P14}, \phw{P15}). 
Improper decisions or outputs from AI tools---especially when uncaught---\qq{can turn out to be very costly. Humans make mistakes too, but when a human makes a mistake, you have various justifications for it, such as someone didn't do their due diligence. If the algorithm makes a mistake, it is harder to explain} (\phw{P14}). 
In the \legal\ context, \leg{P05} describes challenges with AI-generated risk scores for recidivism replacing expert testimony:
\qqq{It used to be that you would work with probation officers and other experts who would say, `Hey! Given this person's age, given their criminal history background, given the age at which they started being in the system, it's more or less likely they will re-offend.' And now, you get a score. For a defense attorney, it's like, `Well, what am I supposed to do with that score?' If this were an expert testifying, I would have the right to cross-examine them. We would have the ability to question their expertise... We are essentially allowing these AI systems to provide expert testimony without any of the protections that we normally afford to expert testimony.}{\leg{P05}}

\med{P09} clarified that decision-makers are \qq{not new to that type of tension where you're confused about something and asking for help} as \medicine\ decision-makers often consult their peers and are used to seeking second opinions (\emph{e.g.}, clinicians frequently come together to discuss complex cases during grand rounds).
Current AI tools for decision-making often only provide one-shot recommendations, which contrasts with established practices of dialogue and clarification that allow decision-makers multiple opportunities to ask questions. Comparatively, participants spoke more positively about generative chatbots, as these tools enable back-and-forth discussion (\jrn{P03}, \med{P09}, \med{P11}).

\paragraph{Faulty beliefs in model capabilities.}
Adoption choices are also shaped by what decision-makers believe AI tools can do---which may not always be accurate. Unclear model limitations can lead to misguided assumptions about model capabilities and adoption benefits.
As one participant noted, AI is sometimes given \qq{an unnecessary and undeserved veneer of neutrality and objectivity,} even though models and their training data are \qq{not inherently objective or neutral, and reflect our own biases} (\leg{P05}). 
The challenge of accurately assessing model quality is further compounded by the rapid proliferation of models across both research and commercial domains. As noted by \med{P08}, models are being \qq{created by all of these different people,} making evaluation difficult. Moreover, limited clinical or professional engagement with modern machine learning methods (\emph{e.g.}, deep learning, neural networks) leaves some decision-makers struggling to \qq{judge what's a really good model} (\med{P08}). As a result, they may base adoption choices on faulty beliefs and overconfidence in AI capabilities.

\subsection{Consequences for the decision-maker}\label{sec:cf:consequences}

When choosing whether to adopt AI tools, decision-makers also weigh potential consequences---including additional \textbf{legal, professional, and reputational liabilities}---against their current workflows.
In particular, decision-makers may consider the risk of potential lawsuits, loss of professional licences, or damaging credibility, citing concerns such as \qq{the reputational risk of being wrong} (\med{P12}). 
Legal liability remains a complex consideration as \qq{laws have still not evolved} to address decisions involving AI models (\med{P09}). In \medicine, \med{P10} emphasizes that \qq{regulating AI models is hard and it all moves at a pace that the FDA does not}. Similarly, in \legal\ contexts, although judges and court systems have begun to issue guidance rules on the use of AI (e.g., the guidance for the use of generative AI from the state bar of California mentioned by \leg{P06}), regulation is still largely \qq{catching up} to actual AI use in practice (\leg{P05}). 

In addition, \textbf{uncertainties about how these consequences apply} may also discourage adoption.
For instance, clinicians in \medicine\ ask: \qqq{Who is liable? Is it the AI? Is it the company or the programmer that made it? That then a health system purchased? Was [the model] written in-house? Is it the doctor who, at the end of the whole process, is ultimately responsible?}{\med{P14}}
Beyond legal liability, there are also concerns about political and professional risks, among other liabilities. \leg{P05} described how judges may be hesitant to rely on AI tools for fear that incorrect decisions could become a \qq{political liability} if they later run for office (\leg{P05}). 
While some AI vendors have attempted to mitigate these concerns by offering indemnification clauses---\emph{e.g.,} compensating law firms for losses caused by chatbot errors---the individual professional often remains ultimately accountable: \qq{to the client, the lawyer is [still] the sole responsible party for any mistakes} (\leg{P06}). As \leg{P05} summarized, adopting AI tools is often seen as being \qq{at your own peril}.

Beyond legal and reputational consequences, decision-makers also consider practical implications for their daily work, such as \textbf{workload changes}.
Many participants mentioned poor integration with existing workflows as a significant barrier to adoption (\med{P10}, \phw{P13}, \phw{P14}). Especially in high-stress fields where burnout is prevalent, any additional effort required to incorporate AI tools can be prohibitive (\med{P09}). 
In contrast, participants expressed more optimism toward AI adoption in ``low-stakes'' scenarios (\emph{i.e.,} contexts where there are fewer potential negative consequences and the impact on workload is more clearly beneficial). 
For example, \med{P11} noted that \qq{administrative simplification and cutting out processes that are low-stakes and do not have risk for bias are a lot easier to contemplate using AI for than having an agent-based model make decisions that impact patients.} Similarly, \med{P08} described how most AI models currently used in practice address \qq{low-hanging fruit and low-stakes problems such as resource utilization or administrative work.} 
In these cases, decision-makers can leverage the processing capability of AI tools \qq{to streamline their processes to make their work a little bit more efficient} (\phw{P15}).

\subsection{Perceived implications for other stakeholders} \label{sec:cf:stakeholders}
When considering AI adoption, decision-makers also weigh the broader impact of AI and the implications for other stakeholders involved in how AI systems are deployed and used.
These stakeholders' goals may diverge from those of the decision-maker. 
One key stakeholder is the \textbf{organization} to which the decision-maker belongs.
For instance, in \journalism, news agencies \qq{obviously have different agendas} from journalists and typically prioritize profits \jrn{P02}, whereas many journalists value producing high-quality content over maximizing output. 
Nonetheless, organizational objectives such as cost reduction and efficiency gains can significantly influence decision-makers' adoption choices.

\textbf{Decision-subjects} (\emph{e.g.,} patients in \medicine\ settings or clients in \legal\ contexts) are another crucial stakeholder group considered by decision-makers. While AI tools can offer benefits to decision-makers and their organizations, participants noted potential harms to decision-subjects. For example, AI systems may fail to identify contextually relevant factors that a human might catch (\leg{P05}), or introduce biases and stereotyping in decision-making (\jrn{P04}, \leg{P06}, \med{P09}, \med{P10}, \med{P12}, \phw{P15}). These risks may make decision-makers hesitant to adopt AI tools, especially when decision-makers perceive risks to vulnerable populations.

\section{Comparative Analysis of AI Adoption}\label{sec:case-study}

We use the factors outlined in Section~\ref{sec:cf} to develop an AI adoption sheet (Figure~\ref{fig:checklist}) that helps analyze decision-makers' adoption choices around adopting AI for specific use cases. We define these use cases as unique contexts where AI supports decision-making or task completion. Importantly, decision-maker adoption is often context-dependent, varying not only across different use cases, but also for similar use cases in different domains.  

To identify use cases of interest, we first coded 34 distinct use cases mentioned by our participants (detailed in the Appendix). 
Most use cases are domain-specific, targeting specific tasks or objectives (\emph{e.g.,} recidivism prediction in \legal\ or \medicine\ diagnostics through radiology reads). 
However, two use cases appeared in three domains each: \textbf{e-discovery} (mentioned by nine participants in \journalism, \legal, \welfare) and \textbf{tailoring communications} (mentioned by four participants in \journalism, \medicine, \welfare).
We now examine both case studies through a cross-domain lens to illustrate how our AI adoption sheet helps systematically elucidate differences in decision-maker adoption.

\begin{figure*}[t]
  \centering
  \begin{subfigure}[b]{0.99\linewidth}
    \centering
    \includegraphics[width=0.99\linewidth]{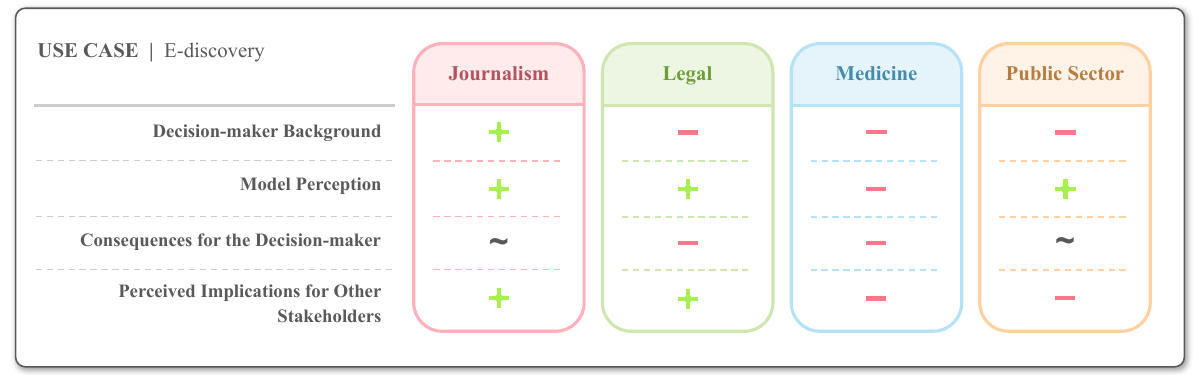}
    \caption{E-Discovery for Research and Analysis}
    \label{fig:ediscovery}
  \end{subfigure}
  \begin{subfigure}[b]{0.99\linewidth}
    \centering
    \includegraphics[width=0.99\linewidth]{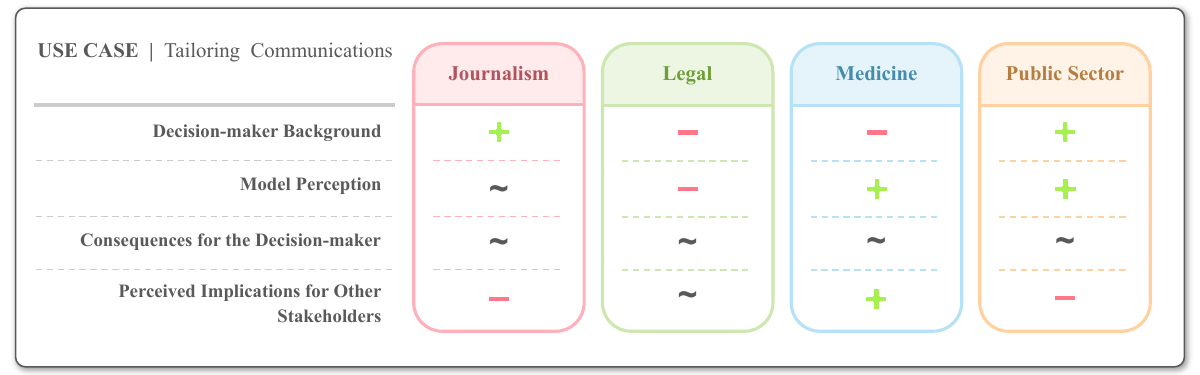}
    \caption{Tailoring Communications to Audiences of Different Backgrounds}
    \label{fig:tailoring}
  \end{subfigure}
  \caption{An overview of two cross-domain case studies analyzed through our AI adoption sheet. For each domain, a green plus sign (+) indicates that a factor is likely to influence decision-makers to adopt AI tools, a red minus sign (--) indicates that a factor is likely to sway decision-makers NOT to adopt AI tools, and a tilde ($\sim$) indicates that the factor likely does not strongly affect a decision-maker's adoption choice.
  }
\end{figure*}

\subsection{E-Discovery for Research and Analysis}\label{sec:case-study:ediscovery}

Decision-makers increasingly use AI tools for \emph{e-discovery}, which refers to gathering, organizing, and analyzing relevant digital data for downstream goals.
For example, journalists investigating a story often need to draw on various sources of data (\emph{e.g.,} interviews, public records)~\citep{stray2021making, broussard2019artificial,quinonez2024new}. 
All of our \journalism\ participants described AI tools for e-discovery to be extremely helpful in streamlining the reporting process (\jrn{P01}, \jrn{P02}, \jrn{P03}, \jrn{P04}). 
In \legal\ contexts, discovery is a long-standing procedure that comprises the initial phase of litigation, requiring attorneys and paralegals to sift through case materials to identify pertinent documents and evidence. 
Recently, AI tools have allowed attorneys to more efficiently identify relevant documents in the increasingly digitalized landscape through e-discovery (\leg{P05}, \leg{P06})~\citep{pai2023exploration, nagineni2024rise, yang2024beyond}. 
\welfare\ decision-makers similarly draw on community-level data to help shape effective policies, and participants described the use of AI tools for the analysis of these data (\phw{P13}, \phw{P15}, \phw{P16})~\citep{shang2024leveraging,yang2024automating}. 
In \medicine\ settings, a key part of clinicians' workflow involves reviewing patient histories to tailor treatment plans and provide the highest quality of care for each patient. 
AI for e-discovery could substantially reduce their workload; however, our \medicine\ participants noted that these types of tools are seldom adopted in practice (\med{P08}, \med{P12}).\footnote{A large-scale survey by PWC studying AI and robotics in healthcare has also not identified the use of AI tools for e-discovery in \medicine~\citep{pwc2017ai}.}
We summarize our analysis in Figure~\ref{fig:ediscovery}, where we illustrate why e-discovery AI tools are more widely adopted by decision-makers in some domains (\emph{e.g.,} in \journalism, \legal, \welfare) compared to others (\emph{i.e.,} in \medicine).

\paragraph{\textbf{Factor 1: Decision-maker background. }}
In fields requiring high specialization, such as \legal, \medicine, and \welfare, decision-makers often have a greater sense of confidence in their own judgment.
Such confidence may make them less inclined to rely on support from AI tools, especially if the tools might lack relevant context (\leg{P05}, \med{P09}). 
For example, an attorney may believe \qq{AI tools have a hard time understanding the nuance and complex context involved} in case documents and thus prefer to go through discovery on their own (\leg{P06}). 
\leg{P05} notes that although AI tools for e-discovery can be extremely beneficial, \qq{there's a trade-off. Being the associate who knew the entire case and read every document was really valuable and obviously, there are times when a machine learning algorithm is not able to correctly predict what would have been relevant whereas a lawyer would have noticed something and pulled that document.}
We also observe this in \medicine, where clinicians may prioritize their professional experience over insights provided by AI (\med{P08}, \med{P12}). 
Even in \welfare, case workers often have extensive professional experience with the communities they serve, allowing them to process more nuanced context than AI tools can provide (\phw{P14}). 
In contrast, decision-makers in \journalism\ may tackle a wide range of topics, including those outside their immediate expertise, which may make them more likely to embrace AI tools for e-discovery (\jrn{P02}, \jrn{P03}).

\paragraph{\textbf{Factor 2: Model perception. }}
E-discovery tools in \journalism\ and \legal\ settings are typically used for researching information as decision-makers believe they can simplify and streamline research processes effectively (\jrn{P03}, \leg{P05}, \leg{P06}, \leg{\citep{reuters2024genai}}). 
Similarly, decision-makers believe e-discovery tools in \welfare\ can be used to support existing data analysis frameworks (\emph{e.g.,} generating spreadsheet formulas).
As \phw{P15} explained: \qq{If I ask an AI tool for a formula to do something in sheets and it gives me an answer that's not true, it's easy for me to know that it's not true because when I just go in, I type in the formula and it doesn't work}. 
In these settings, decision-makers may perceive models to be more trustworthy as it is easier to identify when AI outputs are incorrect. 
In contrast, participants in \medicine\ domains tend to express significant skepticism towards the use of AI tools and algorithms in clinical practice (\med{P10}) as it remains \qq{unclear what their added value is and whether there is really consensus that these models are safe to use}(\med{P08}).
\med{P09} admits it feels \qq{a bit off when you have the human depending on the accuracy of the model and not on the expertise of humans.} 
Especially as there are a multitude of established and validated resources in \medicine\ (\emph{e.g.,} classical risk scores), clinicians may be unlikely to adopt AI tools for e-discovery.

\paragraph{\textbf{Factor 3: Consequences for the decision-maker. }}
The stakes of using e-discovery tools vary across domains.
For example, attorneys may face serious legal repercussions if e-discovery tools cause pertinent information to be missed or lead the attorney to form biased opinions, negatively impacting trial outcomes. 
Improper handling of sensitive data may also have legal ramifications for the attorney considering \qq{chats with AI systems are not confidential} (\leg{P06}) and would \qq{be a breach of duties as a lawyer to protect client confidential information,} violating attorney-client privilege without informed consent (\leg{P05}). 
In \medicine, similar concerns have arisen regarding incorporating e-discovery tools, though the full extent of potential consequences remains unclear. 
As noted by \med{P10}, malpractice liability is a key consideration among clinicians: \qqq{All the clinicians have malpractice insurance. Do the algorithms? Do they need to? I think all of these are hard questions. If I was liable for what happened to my patient and I have very expensive insurance because of that, I don't want to be told what to do unless they're going to take the liability away.}{\med{P10}} 
In high-stakes domains such as \legal\ and \medicine---where risks are more pronounced---decision-makers may be more hesitant to adopt tools that could expose them to additional liability. 
In comparison, mistakes as a result of using AI tools during the research step in both \journalism\ and \welfare\ may only result in a mild reprimand.

\paragraph{\textbf{Factor 4: Perceived implications for other stakeholders. }}
Since e-discovery tools can make decision-makers more efficient, other stakeholders may prefer decision-makers to adopt these tools.
For example, \leg{P05} and \leg{P06} noted that e-discovery tools powered by generative AI, when used appropriately, can help not only lawyers, but also \qq{clients because it can make what lawyers do cheaper} as clients are billed based on the hours spent on their case (\leg{P05}). 
Some law firms may also encourage the usage of these tools so that lawyers can take on more clients.
Similarly, news companies that journalists work for may also urge the use of e-discovery tools to help maximize content output, and thus business profits as well (\jrn{P02}, \jrn{P03}). 
However, in \medicine, while some hospitals may promote the adoption of AI for e-discovery to increase the number of patients that a clinician sees each day, clinicians tend to feel that they must take the patient into account above everything else. 
To this end, \med{P08} emphasized that AI outputs, which oftentimes ``objectively seem to be the best, may subjectively be worse for the patient'' and may cause more harm than good if adopted. 
Analogously in \welfare, case workers often work for non-profits or government agencies that have ``limited capacity to maintain and use more complex models'' (\phw{P14}), which may result in a limited number of tools available to decision-makers.

\subsection{Tailoring Communication to Audiences of Differing Backgrounds}\label{app:sec:case-study-tailoring}

Effective communication is crucial for successful interactions across numerous domains and is often necessary to communicate between individuals from different backgrounds (\emph{e.g.,} educational, cultural, linguistic). 
Tailoring communication to audiences of differing backgrounds is a complex task that has seen an increase in the use of AI tools~\citep{abdelali2023ai, algouzi2023study, ayvazyan2024things, dwivedi2024ai, lee2024artificial}. Our participants also noted the use of AI tools for relevant applications, including language translation (\jrn{P01}, \jrn{P03}), simplifying jargon (\med{P09}, \med{P12}), and even adjusting the tone and style of writing to better align with specific cultural contexts (\phw{P15}, \phw{P16}). 
Despite their potential, these AI tools are not always adopted by decision-makers. While this use case emerged in interviews across \journalism, \medicine, and \welfare, it was absent in \legal.\footnote{It similarly did not appear across surveys in \legal\ conducted by~\citet{reuters2024genai}.} Using our factors, we investigate how AI tools are used to tailor communications across domains (summarized in Figure~\ref{fig:tailoring}). We focus our discussion on Factors 1, 2, and 4 as they have the greatest inter-domain differences. For conciseness, we omit discussion of Factor 3 as inter-domain differences are less distinct.

\paragraph{\textbf{Factor 1: Decision-maker background.} }
Emphasis in some domains prioritizing human-to-human interactions may \emph{discourage} the adoption of AI for tailoring communications (\leg{P05}, \leg{P06}, \med{P08}, \med{P09}). Especially in \legal\ where \qq{lawyers are the world's most highly paid rhetoricians} (\leg{P05}), direct communications with the decision-maker may be highly valued. 
Similarly, AI tools can help streamline communications in \medicine, but \qq{shared decision-making between a doctor and their patient---which has to be customized to what the patient feels is important at that time}---can never be replaced by AI due to the subjective nature of \medicine\ (\med{P08}).
Alternatively, fields such as \welfare\ may be \qq{chronically underfunded, underresourced, understaffed, and overworked}, so a desire to \qq{be more efficient and do more with less} may encourage decision-makers to adopt AI tools to avoid burnout (\phw{P15}).

\paragraph{\textbf{Factor 2: Model perception. }}
Depending on the specific needs of a decision-maker, they may perceive model performance and reliability differently. Although some decision-makers may be inclined to use AI tools for translation as many languages are available, others may be reluctant to adopt such tools because they feel \qq{some languages are not yet working well} (\phw{P16}) or believe that AI generates low-quality outputs (\jrn{P04},~\jrn{\citep{thompson2024shocking}}). 
Participants echo the sentiment found in \leg{\citet{clio2023legal}}, which highlights that \legal\ professionals are interested yet cautious of AI due to the risk of hallucinations (\leg{P05}, \leg{P06}). 
Similarly, in \medicine, both our participants (\med{P09}, \med{P11}) and the surveys by \med{\citet{ama2025augmented}} show that physician sentiment toward enthusiasm for AI translation services trends positive, with a 3\% increase from 2023 to 2024~\citep{ama2025augmented}. 
\phw{P15} explains this with a concrete example: When prompting an AI model for a vaccine analogy, model outputs \qq{tailoring for the Latino and Hispanic communities gave an example of a soccer game} but when prompting for anecdotes \qq{tailored to the Black and African American communities, the model gave me a bulletproof vest analogy which, of course, is hugely problematic.}
However, in a recent repeat of this experiment, \phw{P15} found the model had improved and no longer performed in this problematic way.

\paragraph{\textbf{Factor 4: Perceived implications for other stakeholders. }}
AI tools can enable decision-makers to tailor communications to a variety of backgrounds. 
In \medicine, while \qq{the best clinicians are the ones who personalize their decisions around the patient,} patients must be able to fully understand any health considerations to properly \qq{come up with a shared decision} with their clinician (\med{P08}). 
As such, our participants spoke positively about AI tools that can translate medical text into \qq{a reading format that fits the patient and their education level} (\med{P09}) to \qq{help patients or caregivers when they have to make personal, value-based decisions by turning something of high complexity into something understandable for them} (\med{P12}). 
Patients have even been found to prefer AI responses to clinician-to-patient messages~\med{\citep{kim2024perspectives, tai2024ai}}.
However, when decision-makers believe that AI tools fail to account for varying cultural backgrounds and the impact of harmful speech, they may be less inclined to adopt AI tools. 
This is because the outputs of AI models \qq{become representative of a country and representative of a culture}, and models trained on low-quality data can perpetuate harmful stereotypes (\jrn{P04}). 
Additionally, \phw{P15} notes that AI use can have a \qq{broader impact. If we continue to use language that is harmful to communities, if we use language that is inaccurate, if we use tools that provide information that is misleading, or that turns people off to the information because it is not presented in a compassionate way. I think all of that will have a negative impact down the line} such as \qq{contribute to inequities and contribute to widening disparities} (\phw{P15}). 
\section{Discussion}\label{sec:discussion}

\subsection{Recommendation for better AI adoption}\label{sec:recommendations}

While decision-makers ultimately choose whether to adopt AI tools, their choices are shaped by many other actors—including model developers, policymakers, and researchers. Building on our findings, we offer four recommendations to support responsible AI adoption in practice.

\paragraph{\textbf{(1) Integrate Decision-Maker Perspectives in AI Development.}} We recommend \textit{starting with the decision-maker and considering whether they are likely to adopt an AI tool} before developing and deploying AI tools for specific use cases. We present our AI adoption sheet in Figure~\ref{fig:checklist}, summarizing the four high-level factors that we identified. 
We also provide examples of how this AI adoption sheet can be applied to the aforementioned case studies in the Appendix. 
Prior to deploying new AI tools, we encourage model developers, policymakers, and researchers to utilize this AI adoption sheet to proactively assess how AI tools may be adopted by decision-makers in practice.

\paragraph{\textbf{(2) Bolster Decision-maker Education on AI.}}
Increasing AI literacy can mitigate the influence of decision-maker background and model misperception on adoption choices~(\med{P07}, \med{P08}, \phw{P16}). 
We encourage decision-makers to engage with emerging resources on AI literacy~\citep{ng2021conceptualizing, chiu2024artificial}. 
We also recommend that organizations developing AI tools concurrently invest in literacy initiatives in collaboration with AI researchers and model designers. 
In particular, we encourage future literacy work to focus not only on decision-maker adoption choices, but also on the relationship between adoption and reliance to promote responsible use of AI tools by decision-makers.

\paragraph{\textbf{(3) Develop Infrastructure to Integrate AI into Existing Workflows.}}
Oftent, the additional workload required to use AI tools not integrated into decision-maker workflows may be prohibitive to adoption. Moreover, AI systems often demand substantial resources---such as data storage---that many organizations, particularly those lacking a clear AI adoption strategy, struggle to accommodate (\jrn{P01}, \leg{P06}, \med{P10}, \phw{P13}, \phw{P14}). These challenges typically place an additional burden on the decision-maker, further deterring adoption. We recommend that organizations adopt a more deliberate approach to AI procurement, ensuring they have the necessary infrastructure in place. To support decision-maker adoption, deployed tools should align with existing workflows rather than requiring substantial procedural changes~\citep{silva2025procuring}.

\paragraph{\textbf{(4) Provide Resources to Decision-makers for Navigating Applicable Policies.}}
The consequences decision-makers must weigh when adopting AI tools remain poorly defined (\med{P10}, \med{P11}). Our interviews suggest that the current regulatory landscape is inadequate, with one participant noting a \qq{lack [of] the right regulatory framework} to ensure responsible AI use (\leg{P05}). 
The reactive nature of AI policy has left decision-makers struggling to \qq{navigate the patchwork regulatory approach that currently exists} (\leg{P05}). Policymakers should interface with existing research on AI governance~\citep{schiff2020s, stix2021actionable, schiff2022education} and collaborate with researchers, model designers, and decision-makers to craft more coherent and proactive AI policies.
However, until such policies are implemented, it is essential that organizations offer practical resources to their members to support navigation of the existing regulatory environment.

\subsection{Comparison of findings to prior work}\label{sec:discussion:comparison}

\paragraph{Comparison to organizational factors.} As highlighted in Section~\ref{sec:rw:org}, researchers have extensively studied AI adoption at the organizational level, identifying factors such as issues with data quality and availability, lack of data governance or regulation, and goals related to cost or time efficiency. 
These organizational-level factors parallel some of our decision-maker-centric factors---for instance, organizational concerns about data quality or availability mirror decision-maker concerns about model performance (\emph{Model perception}). 
However, prior work studying organizational adoption tends to overlook the nuanced and multifaceted consequences considered by decision-makers, which include time trade-offs, various types of liabilities, and implications for other stakeholders that may not align with organizational priorities. 
Although organizational factors touch on some of these issues, they typically focus narrowly on organizational goals. 
Examining adoption at both the decision-maker and organizational levels highlights overlapping factors that are important to consider when evaluating AI tools. 
However, considering AI adoption from the perspective of decision-makers may offer a more comprehensive understanding of the factors shaping real-world adoption.

\paragraph{Comparison to previously identified decision-maker factors.} While prior work has rarely studied adoption at the level of the individual decision-maker, our findings align with and extend insights from two prior works~\citep{fang2023ai,hameed2023breaking}. 
\citet{fang2023ai} emphasizes that distinctions based on professional roles---analogous to \emph{Decision-maker background}---are important to understanding decision-maker adoption in \legal. However, their findings focus solely on how occupational roles shape adoption attitudes and do not cover any other factors.
\citet{hameed2023breaking} validates seven predefined constructs to influence adoption behaviors (performance expectancy, effort expectancy, personal innovativeness in information technology, task complexity, technological characteristics, initial trust, and social influence) that also align with our identified factors. 
However, while these studies find factors to always be significant, our case studies in Section~\ref{sec:case-study} show that these factors vary in importance across domains and use cases. By offering broader categories with more granular prompts, our AI adoption sheet helps organize and simplify these constructs, encouraging more systematic future research into what shapes AI adoption choices. Together, these contributions provide a structured, practical tool for future research and practice.

\subsection{Limitations and Future Work}
\label{sec:discussion:limitations}

Our study has several limitations, which motivate directions for future work.
We acknowledge that our participant sample size and the set of domains are limited and do not capture the full spectrum of real-world applications. Additionally, our participants were all based in the United States, which may limit the generalizability of our findings to other domains and contexts. As such, the four key factors we identified in Section~\ref{sec:cf} may not be exhaustive. We encourage future work to build on our methods and evaluate a larger sample of decision-makers in more domains. Additionally, as our recruitment protocol required that study participants have prior experience with AI tools in their domain, our interviewees may be biased towards the adoption of AI tools in practice. Participants may have also had varying degrees of comfort when discussing some use cases and factors over others, which may have shaped the responses we received. As our interviews were open-ended, discussions primarily focused on an \emph{individual} decision-maker, which is consistent with most prior work. However, we encourage future work to also consider factors influencing adoption by groups of decision-makers.
Further, we acknowledge that decision-makers' perceptions and the capabilities of AI models are rapidly evolving. Our interviews, conducted between October 2023 and October 2024, reflect perceptions and model capabilities relevant during this time frame. 
Finally, we focused on \emph{inter}-domain differences; further study is needed to explore \emph{intra}-domain differences between decision-makers and how they influence AI adoption.
\section{Conclusion}

AI adoption depends on the decision-maker actively incorporating AI tools into their workflows.
To identify common factors across domains that influence decision-makers' use of AI, we conducted 16 semi-structured interviews with decision-makers across four domains.
Through our cross-domain analysis, we identify four key factors that impact decision-maker adoption.
We apply this sheet to two case studies of real-world use cases, illustrating why AI-powered tools are more widely adopted by decision-makers in some domains and not others.
We believe that the AI adoption sheet has many uses; it can support researchers considering the adoption of AI in new domains, model developers creating new AI tools with the end user in mind, and organizations identifying appropriate AI tools suited for decision-maker adoption. 
In closing, we hope that this work not only encourages further analyses focused on the decision-maker but also inspires future research to look beyond a single use case or domain.

\bibliographystyle{ACM-Reference-Format}
\bibliography{references}

\appendix

\section{IRB-approved Protocol}
\subsection{Email Solicitation}
Dear [...],

We hope our email finds you well. We are a group of researchers from Carnegie Mellon University, conducting a research study on Human-AI (H-AI) partnerships in practice. Based on the publicly available information, we have identified you as an individual with relevant domain expertise with experience working with AI in our application domains of interest (journalism, social computing, medicine, justice, welfare). We are writing to invite you to participate in a virtual interview.

\textbf{Study purpose:} We have observed that H-AI partnerships look different in various domains and as such are interested in studying how different factors may affect these partnerships and the outcomes of these partnerships. As such, the purpose of the study is to obtain the opinions of domain experts across several important domains (e.g., medicine, justice, etc.) on applications of Artificial Intelligence (AI) and existing H-AI partnerships in their field in order to explore how various aspects of these partnerships may augment or diminish their use in practice. We consider H-AI interactions, collaboration between humans and AI models, and AI-assisted decision-making as forms of H-AI partnership. 

\textbf{Participation qualifications:} In order to participate, you must be 18 years or older. Since this study seeks to better understand H-AI partnerships in practice, we are looking to interview individuals who are either:

\begin{itemize}
    \item[(1)] domain experts with experience around human-AI partnerships in real-world applications of AI
    \item[(2)] researchers who have worked closely with domain experts.
\end{itemize}

Interview questions will cover your level of familiarity with applications of AI in your domain—how your field benefits from AI models, what potential harms AI models present, what ideal H-AI partnerships look like to you, etc.—as well as your perceptions of existing research on H-AI partnerships within your domain. If you feel as if your expertise does not enable you to answer these questions comfortably, we thank you for your time and consideration. If you can think of other prominent individuals in the field who you believe would be a good fit for our study, we would appreciate it if you could forward this email to them or share their contact information with us.

\textbf{Study details:} We expect the full study to take less than 60 minutes to provide detailed responses, but all questions are optional to respond to. Participation is voluntary and you will be compensated \$100 for your participation in a one-on-one interview. Please see the consent form attached. If you wish to proceed after reading the consent form, please schedule an interview here: [\texttt{link to scheduling form}]

If you have any questions or concerns at any point, please don’t hesitate to reply to this email to get in touch with the principal investigator (rebeccay@andrew.cmu.edu). 

Thank you for your consideration! Again, if you can think of other prominent individuals in the field who you believe would be a good fit for our study, we would greatly appreciate it if you could forward this email to them or share their contact information with us.

Best regards,

Rebecca Yu (on behalf of the research team)

Valerie Chen

Hoda Heidari

Ameet Talwalkar

\subsection{Interview Protocol}
Interviews were semi-structured, meaning that we started with a set of predetermined interview questions, but also asked additional questions to follow up on participants' responses and explore emerging themes~\citep{adams2015conducting, adeoye2021research}. Our interview protocol, including predetermined interview questions, is as follows:

\begin{itemize}
    \item Interviewer instructions
    \item Consent
    \begin{itemize}
        \item (Before recording) Can you confirm that you have received and read through the consent document?
        \item (On recording) Can you confirm that the consent information has been provided to you, that you agree to participate, and that you agree to have the session recorded?
        \item Please remember not to reveal any information you don’t want us to include in our study. If you do say anything you would like us to exclude, please let us know.
    \end{itemize}
    \item Background information
    \begin{itemize}
        \item What is your area of expertise?
        \begin{itemize}
            \item What is your current role?
            \item How many years of domain experience do you have?
        \end{itemize}
        \item What is your level of familiarity with Artificial Intelligence (AI) or Machine Learning (ML)?
        \begin{itemize}
            \item How many years of experience with AI do you have?
        \end{itemize}
        \item Within your domain, how much interaction do you have with AI/ML models?
    \end{itemize}
    \item Applicaiions of AI
    \begin{itemize}
        \item Based on your prior experiences and use cases you are familiar with, what is the use of AI in your domain?
        \item Do humans interact with these AI systems? We’re particularly interested in use cases where humans work closely with AI models in order to make decisions.
        \item Can you elaborate on some of the interactions between humans and AI systems?
        \item Why do you think current interactions are designed as such? What were the main considerations?
        \item What should be the main considerations for setting up H-AI interactions for [insert example they talked about]?
    \end{itemize}
    \item Conclusion
    \begin{itemize}
        \item Do you have any questions for us?
        \item If there’s anything from this interview you would like us to not include, please let us know.
        \item We’ll be reaching out to you again via email after all of our interviews have concluded with regard to reimbursement. Thank you so much for your time!
    \end{itemize}
\end{itemize}

\section{Factors for Organizational Adoption Identified in Prior Work}
Below, we provide a brief survey on the factors identified for organizational adoption:
\begin{itemize}
    \item \citet{cubric2020drivers} --- job security; reusability of model; increased dependence on non-humans; lack of knowledge; trust; safety; availabtility of data; and multiple stakeholders' perspectives.
    \item \citet{harvey2021regulatory} --- data quality; regulatory paradigm and existing pathways; workflow patterns.
    \item \citet{regona2022opportunities} --- reduction of time spent on repetitive tasks; accuracy of events, risks, and cost forecasting; issues with data acquisition and retention.
    \item \citet{bedue2022can} --- ``ability'': system and service quality, access to knowlege (including employee training), transparency, explinability, and reliability, data quality; ``integrity'': standards, guidelines, and government regulation; ``benevolence'': social responsibility, ethical guidelines, data bias, cultural guidelines, sustainability.
    \item \citet{campion2022overcoming} --- resistance to sharing data (privacy and security concerns); insufficient understanding of required data available; lack of alignment between project interests and expectations; engagement across organizational hierarchy.
    \item \citet{madan2023ai} --- cost; alternative novel solutions available; expected benefits; organizational risk taking culture; AI lieracy; data-oriented culture; company leadership; data governance maturity; ecosystem of cpartners and experts; bureauracy and centralized decision-making; data quality; data availability; digital infrastructure.
    \item \citet{merhi2023evaluation} --- high cost of AI; lack of technical expertise; AI integration complexity; IT infrastructure; accessability of a scalable and flexible system; management support; resistance to changing from the status quo; organizatioanl culture; selection of vendors; lack of visibility on benefits of AI; availability of a project champion; ambiguous strategic vision; security and confidentiality; ethics issues; responsibility and accountability; data governance issues; low data quality; insufficient quantity of data.
    \item \citet{uren2023technology} --- cost optimization; technology readiness and AI maturity; disruption of AI adoption; data governance processes; data readiness.
    \item \citet{neumann2024exploring} --- impact on business practices; AI maturity level; alternative solutions available; understandability and explainability of AI solution; fit into existing workflow; organizational innovation culture; intrinsic motivation of project members; data infrastructure and strategy.
    \item \citet{polisetty2024determines} --- cost savings; AI compatibility; improvements to accuracy, speed, and consistency with better business outcomes; perceived trust by employees; data governance; data quality.
\end{itemize}

\section{AI Adoption Sheet: E-Discovery} \label{app:ediscovery-sheet}

We provide an example of how the AI adoption sheet can be utilized to evaluate the use case of e-discovery across four domains in Figure~\ref{fig:ediscovery-sheet}.

\begin{figure*}[htp]
    \centering
    \textbf{Decision-maker adoption for E-Discovery}\\~\\
    \begin{subfigure}[t]{0.45\textwidth}
    \fcolorbox{black}{white}{
    \begin{minipage}[t]{0.9\columnwidth}
        \centering \textbf{Journalism (\journalism)}

        \begin{itemize}[left=5pt]
            \item \textbf{Decision-maker background.} 
            \begin{itemize}[left=3pt]
                \item \textbf{Professional experience.} Decision-makers are comfortable using AI tools for e-discovery to augment journalism.
            \end{itemize}
            \item \textbf{Perceptions of the AI model.} 
            \begin{itemize}[left=3pt]
                \item \textbf{Model transparency and trustworthiness.} Outputs can be easily verified/validated.
                \item \textbf{Belief in model capabilities and adoption benefits.} Adopting AI tools be helpful in streamlining the journalism process.
            \end{itemize}
            \item \textbf{Consequences for the decision-maker.} 
            \begin{itemize}[left=3pt]
                \item \textbf{Liabilities.} With due diligence, there should be few legal, professional, or social liabilities.
                \item \textbf{Uncertainty regarding consequences.} Potential consequences are uncertain due to a lack of regulation.
            \end{itemize}
            \item \textbf{Implications for other stakeholders.} 
            \begin{itemize}[left=3pt]
                \item \textbf{Organizations.}  \journalism\ organizations will prioritize a high quantity of articles, which adopting an AI e-discovery tool will help with.
            \end{itemize}
        \end{itemize}
    \end{minipage}
    }
    \end{subfigure}
    ~
    \begin{subfigure}[t]{0.45\textwidth}
    \fcolorbox{black}{white}{
    \begin{minipage}[t]{0.9\columnwidth}
        \centering \textbf{Public Sector (\welfare)}

        \begin{itemize}[left=5pt]
            \item \textbf{Decision-maker background.} 
            \begin{itemize}[left=3pt]
                \item \textbf{Professional experience.} \welfare\ caseworkers are likely to prioritize their expertise over AI outputs.
            \end{itemize}
            \item \textbf{Perceptions of the AI model.} 
            \begin{itemize}[left=3pt]
                \item \textbf{Model flaws.} Some concerns about hallucinations and misinformation.
                \item \textbf{Model transparency and trustworthiness.} Outputs are easy to fact check.
                \item \textbf{Belief in model capabilities and adoption benefits.} Added value when used for data processing tasks---including time optimization.
            \end{itemize}
            \item \textbf{Consequences for the decision-maker.}
            \begin{itemize}[left=3pt]
                \item \textbf{Uncertainty regarding consequences.} Unclear liabilities for the decision-maker.
            \end{itemize}
            \item \textbf{Implications for other stakeholders.}
            \begin{itemize}[left=3pt]
                \item \textbf{Organizations.} \welfare\ organizations may lack the infrastructure to support AI adoption.
                \item \textbf{Decision-subjects.} Biased, discriminatory, or stereotyping outputs may significantly harm decision-subjects.
            \end{itemize}
        \end{itemize}
    \end{minipage}
    } 
    \end{subfigure}
    \\~\\~\\
    \begin{subfigure}[t]{0.45\textwidth}
    \fcolorbox{black}{white}{
    \begin{minipage}[t]{0.9\columnwidth}
        \centering \textbf{Law (\legal)}

        \begin{itemize}[left=5pt]
            \item \textbf{Decision-maker background.} 
            \begin{itemize}[left=3pt]
                \item \textbf{Professional experience.} Lawyers and other decision-makers in \legal\ are likely to prioritize their expertise over AI outputs. 
            \end{itemize}
            \item \textbf{Perceptions of the AI model.} 
            \begin{itemize}[left=3pt]
                \item \textbf{Belief in model capabilities and adoption benefits.} Models are trained specifically on \legal\ data. 
            \end{itemize}
            \item \textbf{Consequences for the decision-maker.} 
            \begin{itemize}[left=3pt]
                \item \textbf{Liabilities.} Likely significant legal and professional liabilities.
                \item \textbf{Uncertainty regarding consequences.} Potential indemnification clauses, though these do not absolve the decision-maker of liability.
                \item \textbf{Workload changes.} Potential time optimization from AI adoption may be offset by the time required to fact-check.  
            \end{itemize}
            \item \textbf{Implications for other stakeholders.} 
            \begin{itemize}[left=3pt]
                \item \textbf{Organizations.} \legal\ firms may advocate for AI adoption for time efficiency.
                \item \textbf{Decision-subjects.} Some clients may want AI tools for time efficiency and consequently reduced cost, while others prefer no use of AI.
            \end{itemize}
        \end{itemize}
    \end{minipage}
    } 
    \end{subfigure}
    ~
    \begin{subfigure}[t]{0.45\textwidth}
    \fcolorbox{black}{white}{
    \begin{minipage}[t]{0.9\columnwidth}
        \centering \textbf{Medicine (\medicine)}

        \begin{itemize}[left=5pt]
            \item \textbf{Decision-maker background.} 
            \begin{itemize}[left=3pt]
                \item \textbf{Professional experience.} Clinicians are likely to prioritize their expertise over AI outputs. Healthcare treatment requires subjective treatment plan (human interaction), not objective (model output).
            \end{itemize}
            \item \textbf{Perceptions of the AI model.} 
            \begin{itemize}[left=3pt]
                \item \textbf{Model transparency and trustworthiness.} Models are not transparent or as well-validated when compared to existing models.
                \item \textbf{Belief in model capabilities and adoption benefits.} Time optimization, though the benefits are uncertain given the time necessary to fact-check. 
            \end{itemize}
            \item \textbf{Consequences for the decision-maker.} 
            \begin{itemize}[left=3pt]
                \item \textbf{Liabilities.} Decision-maker is at risk for significant legal and professional liabilities if adverse outcome.
                \item \textbf{Uncertainty regarding consequences.} No clear guidelines regarding who is liable.
            \end{itemize}
            \item \textbf{Implications for other stakeholders.} 
            \begin{itemize}[left=3pt]
                \item \textbf{Organizations.} \medicine\ hospitals may advocate for AI adoption for time efficiency.
                \item \textbf{Decision-subjects.} Biased, discriminatory, or stereotyping outputs may significantly harm decision-subjects.
            \end{itemize}
        \end{itemize}
    \end{minipage}
    } 
    \end{subfigure}
    \caption{For each domain, we show how one could proactively work through the AI adoption sheet for the e-discovery case study to understand decision-maker adoption. For brevity, we only include key subfactors for each domain.}
    \label{fig:ediscovery-sheet}
\end{figure*}
\section{Taxonomizing Use Cases}
\label{app:sec:use-case-taxonomy}

We coded our interviews for use cases mentioned by participants. These use cases were then categorized as utilizing predictive or generative AI models.
Predictive AI use cases mentioned by participants are described in Tables~\ref{app:ucs-predai1} and~\ref{app:ucs-predai2}. Generative AI use cases mentioned by participants are described in Tables~\ref{app:ucs-genai1} and~\ref{app:ucs-genai2}.

\begin{table*}[!htp]
    \centering
    \fontsize{9}{9}
    \begin{tabular}{p{0.2\textwidth} p{0.6\textwidth} p{0.1\textwidth}}
    \toprule 
    \textbf{Use Case}   &
    \textbf{Description}    &
    \textbf{Mentions}   \\

    \midrule
    
      \raggedright\textbf{UC01} \newline Medical risk stratification 
    & \raggedright Derive risk scores from logistic regression models for patients across various applications (e.g., Wells' Criteria for pulmonary embolisms, qSOFA score for sepsis) to identifiy high-risk individuals for triaging or diagnosis. 
    & \med{P08},~\med{P10} \\

    \midrule

      \raggedright\textbf{UC02} \newline Recidivism prediction
    & \raggedright Generate risk scores that are provided to judges attempting to identify ideal sentencing terms to result in just \qq{enough time in prison that is necessary to serve deterrent purposes and rehabilitative purposes} (\leg{P05}).
    & \leg{P05} \\

    \midrule

      \raggedright\textbf{UC03} \newline Risk stratification to identify at-risk individuals
    & \raggedright Identify children at a high-risk of harm or at risk of returning to the foster care system
    & \phw{P14} \\

    \midrule

      \raggedright\textbf{UC04} \newline Targeting of welfare services
    & \raggedright Determine risk scores for individuals and/or communities to identifiy high risk for optimized allocation of welfare services
    & \phw{P13},~\phw{P14} \\

    \midrule

      \raggedright\textbf{UC05} \newline Assisting medical imaging reads by experts
    & \raggedright Provide imaging diagnosis to an expert decision-maker to aid their diagnosis process. Decision-makers may review the model's read prior and simply review results or they may review the model's read after they make their own initial diagnosis prior to their final diagnosis.
    & \med{P07},~\med{P08},~\med{P09} \\

    \midrule

      \raggedright\textbf{UC06} \newline Initial radiology reads for non-experts
    & \raggedright Provide radiology read given medical imaging directly to non-expert decision makers (e.g., clinicians that are not trained in radiology). Reads are reviewed by expert radiolgists within 48 hours (or 2 hours in urgent cases) who may issue addendums as needed.
    & \med{P07} \\

    \midrule

    \raggedright\textbf{UC07} \newline ``Gut-check'' medical diagnoses
    & \raggedright AI identification of easy diagnoses (e.g., obvious cases of pneumoonia) as a ``just in case you missed it'' backup
    & \med{P08},~\med{P10} \\

    \midrule

      \raggedright\textbf{UC08} \newline Hospital case scheduling
    & \raggedright Generate heuristic-based risk scores for each case in order to optimize scheduling of elective and emergency medical cases for effective time and resource allocation
    & \med{P10} \\

    \midrule

    \raggedright\textbf{UC09} \newline Visualizing risk for informed resource-allocation
    & \raggedright Generate risk scores to identify high risk areas that are presented to decision-makers to help with informed resource-allocation decisions
    & \phw{P13} \\

    \midrule

    \raggedright\textbf{UC10} \newline Prioritizing public health check-in calls 
    & \raggedright Generate heuristic-based risk scores for each case in order to optimize calling schedule for medical and/or public health experts to identify at-risk individuals or otherwise urgent scenarios where experts should focus initial calling efforts to optimize outcomes given limited resources
    & \phw{P13} \\

    \midrule

    \raggedright\textbf{UC11} \newline Fraud detection in medical insurance claims
    & \raggedright Parse insurance claims and records to identify suspicious claims and flag providers who may be overbilling patients for further review by an expert decision-maker.
    & \med{P11} \\

    \midrule

    \raggedright\textbf{UC12} \newline Medical image flagging for prioritized clinical review
    & \raggedright Identify and flag medical imaging with life threatening findings for immediate review by a clinician, specifically screening out low-risk images to help decision-makers focus their efforts more effectively.
    & \med{P11} \\

    \bottomrule
    \end{tabular}
    \caption{Summary of predictive AI use cases.}
    \label{app:ucs-predai1}
\end{table*}

\begin{table*}[!htp]
    \centering
    \fontsize{9}{9}
    \begin{tabular}{p{0.2\textwidth} p{0.6\textwidth} p{0.1\textwidth}}
    \toprule 
    \textbf{Use Case}   &
    \textbf{Description}    &
    \textbf{Mentions}   \\

    \midrule

    \raggedright\textbf{UC13} \newline Descriptive analysis of medical images
    & \raggedright Automating tasks such as counting mitotic figures or calculating the amount of calcium in an image
    & \med{P11} \\

    \midrule

    \raggedright\textbf{UC14} \newline Automatic hours tracking and billing
    & \raggedright Automatically track decision-maker (typically an attorney) time allocation when they are working on a client's case. This time allocation is then reviewed and finalized by the decision-maker.
    & \leg{P07} \\

    \midrule

    \raggedright\textbf{UC15} \newline Predictive analytics for data-informed policymaking
    & \raggedright Generate rule-based risk scores that are then benchmarked to consistently apply policy (i.e., automatically determine insurance costs for policyholders) across all scenarios with varying case-by-case decision-maker involvement.
    & \med{P11} \\

    \midrule

    \raggedright\textbf{UC16} \newline Feedback to capture better quality medical images
    & \raggedright Imaging technicians are provided feedback from the model to attain better imaging quality
    & \med{P07}, \med{P09} \\

    \midrule

    \raggedright\textbf{UC17} \newline Improved training in pathology
    & \raggedright Identify the human learning process to improve teaching (and the learning experience of) new decision-makers
    & \med{P07},~\med{P09},~\med{P11} \\

    \bottomrule
    \end{tabular}
    \caption{Summary of predictive AI use cases (cont.).}
    \label{app:ucs-predai2}
\end{table*}

\begin{table*}[!htp]
    \centering
    \fontsize{10}{10}
    \begin{tabular}{p{0.2\textwidth} p{0.6\textwidth} p{0.1\textwidth}}
    \toprule 
    \textbf{Use Case}   &
    \textbf{Description}    &
    \textbf{Mentions}   \\

    \midrule

    \raggedright\textbf{UC18} \newline Medical chart summarization
    & \raggedright Summarize a patient's medical history to provide the clinician a concise update on information relevant to the patient's visit
    & \med{P08} \\

    \midrule

    \raggedright\textbf{UC19} \newline General information summarization
    & \raggedright Decision-makers use the AI model to obtain summarized information about a new, unfamiliar topic to get a general understanding of the topic.
    & \jrn{P02},~\med{P11} \\

    \midrule

    \raggedright\textbf{UC20} \newline Judicial behavior predictive analytics
    & \raggedright Identify trends for a particular judge based on previous decisions to help lawyers identify the most effective way to build into their legal case.
    & \leg{P05} \\

    \midrule

    \raggedright\textbf{UC21} \newline E-discovery, data management, and data analysis
    & \raggedright Quickly gather, organize, and summarize content about a specific topic to streamline processes (e.g., news production, finding legal precedent, analyzing survey data) for decision-makers.
    & {\raggedright}{\jrn{P01},~\jrn{P02},~\jrn{P03}, \jrn{P04},~\leg{P05},~\leg{P06}, \phw{P13},~\phw{P15},~\phw{P16}} \\

    \midrule

    \raggedright\textbf{UC22} \newline Idea generation for writing
    & \raggedright Given a topic by the decision-maker, identify possible ideas to discuss (e.g., interesting sub-topics, potential issues and implications). The model is used by the decision-maker as a brainstorming tool.
    & \jrn{P02} \\

    \midrule

    \raggedright\textbf{UC23} \newline Outline generation for writing
    & \raggedright Outline a potential news article structure for a journalist to iterate off of based on a given topic and relevant sources
    & \jrn{P02} \\

    \midrule

    \raggedright\textbf{UC24} \newline Content generation
    & \raggedright Draft content (e.g., news article, social media post) a decision-maker to review based on a given topic and relevant sources
    & \jrn{P03} \\

    \midrule

    \raggedright\textbf{UC25} \newline Legal case drafting
    & \raggedright Given a legal case, create an initial draft of legal correspondences, etc. when provided relevant evidence, precedents, judicial opinions. Drafts are reviewed and information is verified by the decision-maker.
    & \leg{P05},~\med{P06} \\

    \midrule

    \raggedright\textbf{UC26} \newline Medical visit note drafting
    & \raggedright Draft medical visit notes (e.g., medical history, symptoms, diagnosis, discharge summaries for continuation of care) from ambient listening during clinical interactions that are then reviwed by a clinician.
    & \med{P08}, \med{P09} \\

    \midrule

    \raggedright\textbf{UC27} \newline Insurance authorization drafting
    & \raggedright Draft insurance authorization documentation based on patient information for clinicians to review
    & \med{P09} \\

    \midrule

    \raggedright\textbf{UC28} \newline Radiology assessment drafting
    & \raggedright Draft an initial radiology report given medical imaging and a radiologist's diagnosis that is then reviewed and verified by the radiologist. Some reports may focus on drafting patient-friendly reports while others focus on recommendation or simple result reporting
    & \med{P07} \\

    \midrule

    \raggedright\textbf{UC29} \newline Healthcare document coding for insurance
    & \raggedright Identify codes and draft coding documents from a clinician's medical notes needed to facilitate communication with insurance companies
    &  \med{P08},~\med{P11}\\

    \midrule

    \raggedright\textbf{UC30} \newline Improving data visualization
    & \raggedright Suggest appropriate visualizations for quantitative data, provide code or formulas to generate visualizations, or design assets to improve presentations
    & \phw{P15} \\

    \bottomrule
    \end{tabular}   
    \caption{Summary of generative AI use cases.}
    \label{app:ucs-genai1}
\end{table*}

\begin{table*}[!htp]
    \centering
    \fontsize{10}{10}
    \begin{tabular}{p{0.2\textwidth} p{0.6\textwidth} p{0.1\textwidth}}
    \toprule 
    \textbf{Use Case}   &
    \textbf{Description}    &
    \textbf{Mentions}   \\

    \midrule

    \raggedright\textbf{UC31} \newline Tailoring communication to specific backgrounds
    & \raggedright Provide alternate communications with reduced jargon and/or complexity to accommodate and facilitate conversations with individuals from different backgrounds. For example, AI tools can summarize patient-physician interactions with reduced jargon in a \qq{reading format that fits the patient} \med{P09}. 
    & \jrn{P03},~\med{P09},~\med{P12}, \phw{P15} \\

    \midrule

    \raggedright\textbf{UC32} \newline Meeting notes summarization
    & \raggedright Identify key points in meeting notes (e.g., corporate meetings) for concise summarization of meeting notes
    & \med{P11} \\
    
    \midrule

    \raggedright\textbf{UC33} \newline Copyediting decision-maker text
    & \raggedright Review and edit decision-maker writing (e.g., grammar, structure)
    & \jrn{P03},~\jrn{P04},~\leg{P05} \\

    \midrule

    \raggedright\textbf{UC34} \newline Content moderation facilitation
    & \raggedright Flag content on social media or other sources for potential misinformation or disinformation requiring further moderation by a decision-maker. For example, some models may be used to used to \qq{flag egregious tweets, [including] potentially misleading claims that might go viral} and every piece of content that is flagged is consequently looked at by a human (\jrn{P01}). 
    & \jrn{P01},~\jrn{P02},~\jrn{P04}, \phw{P15} \\

    \bottomrule
    \end{tabular}   
    \caption{Summary of generative AI use cases (cont.).}
    \label{app:ucs-genai2}
\end{table*}

\end{document}